\documentclass[useAMS,usenatbib,aas_macros]{mn2e}

\usepackage{lscape}
\usepackage{graphicx}
\usepackage{colortbl}
\usepackage{amssymb}
\usepackage{amsmath}
\usepackage{natbib}
\usepackage{times}
\usepackage{aas_macros}
\usepackage{threeparttable}

\newcommand{\msun}{~\mathrm{M}_{\odot}}
\newcommand{\zsun}{~\mathrm{Z}_{\odot}}
\usepackage{color}

\def\simpropto{\lower.2ex\hbox{$\; \buildrel \propto \over \sim \;$}}
\def\ltsim{\lower.5ex\hbox{$\; \buildrel < \over \sim \;$}}
\def\gtsim{\lower.5ex\hbox{$\; \buildrel > \over \sim \;$}}

\begin{document}
\title[CR7 powered by DCBH]{CR7 as powered by a direct collapse black hole}
\title[Detecting DCBHs: Is CR7 the one?]{Detecting Direct Collapse Black Holes: making the case for CR7}
\author[B. Agarwal, et al.]{Bhaskar Agarwal$^1$\thanks{E-mail:
bhaskar.agarwal@yale.edu}, Jarrett L. Johnson$^2$, Erik Zackrisson$^3$, Ivo Labbe$^4$, 
\newauthor Frank C. van den Bosch$^1$, Priyamvada Natarajan$^1$, Sadegh Khochfar$^5$\\
$^1$Department of Astronomy, 52 Hillhouse Avenue, Steinbach Hall, Yale University, New Haven, CT 06511, USA\\
$^2$ X Theoretical Division, Los Alamos National Laboratory, Los Alamos, NM 87545, USA\\
$^3$ Department of Physics and Astronomy, Uppsala University, Box 515, SE-751 20 Uppsala, Sweden\\
$^4$ Leiden Observatory, Leiden University, NL-2300 RA Leiden, The Netherlands\\
$^5$Institute for Astronomy, University of Edinburgh, Royal Observatory, Edinburgh, EH9 3HJ\\}

\date{00 Jun 2014}
\pagerange{\pageref{firstpage}--\pageref{lastpage}} \pubyear{0000}
\maketitle

\label{firstpage}

\begin{abstract}

We propose that one of the sources in the recently detected system CR7 by Sobral et al. (2015) through spectro-photometric measurements at $z = 6.6$ harbors a direct collapse blackhole (DCBH). We argue that the LW radiation field required for direct collapse in source A is provided by sources B and C. By tracing the LW production history and star formation rate over cosmic time for the halo hosting CR7 in a $\Lambda$CDM universe, we demonstrate that a DCBH could have formed at $z\sim 20$. The spectrum of source A is well fit by nebular emission from primordial gas around a BH with MBH $\sim 4.4 \times 10^6 \msun$ accreting at a 40\% of the Eddington rate, which strongly supports our interpretation of the data. Combining these lines of evidence, we argue that CR7 might well be the first DCBH candidate.

\end{abstract}

\begin{keywords}
Black holes:  direct collapse -- accretion: DCBH -- cosmology: theory of DCBH
\end{keywords}

\section{Introduction}


Observations of the first of $z>6$ quasars \citep{Fan:2003p40,Mortlock:2011p447,Venemans:2013p3633,2015Natur.518..512W} pose a conundrum for the existence of supermassive black holes (SMBH) with $M_{\bullet}\sim 10^9 \msun$. There is growing consensus that an alternate seeding mechanism, beyond their origin as stellar mass remnants from the first stars, may best explain these high redshift SMBHs.  Whether these are seeded by the $M_{\bullet} \sim 10^2-10^3 \msun $ remnants of the first (Population III) stars forming from metal free gas \citep[e.g.][]{VolonteriRees2006, Alvarez:2009p778} , intermediate mass black holes $M_{\bullet} \sim 10^3 - 10^4 \msun $ resulting from the runaway collapse of dense primordial star clusters \citep{1978MNRAS.185..847B,2004Natur.428..724P}, or massive seeds that resulted from the direct collapse of metal free gas into $M_{\bullet} \sim 10^4 - 10^5 \msun$ black holes, is still an open question \citep[see e.g.][]{Eisenstein:1995p870,Oh:2002p836,Bromm:2003p22,Koushiappas:2004p871,Lodato:2006p375}. In order to disentangle these three distinct seeding mechanisms, we need to find evidence for the formation of a seed BH and to observe its growth during the early stages rather than just the SMBH that it grew into later on.

The recent observation of the brightest Lyman--$\alpha$ (Ly-$\alpha$) emitter found at $z\sim 6.6$ in the COSMOS field \citep[M15 hereafter]{Matthee15a} has opened up discussion about its components \citep[S15 and P15 hereafter]{Sobral15a,Pallotinni15a,TilmanCR7,DijkstraCR7} and their properties. Hubble Space Telescope (HST) imaging presented by S15 reveals that CR7 is in fact made up of 3 distinct clumps (A, B \& C) that appear to be $\sim$5 kpc apart (projected distance). The components B \& C appear to be evolved galaxies, whereas the bulk of the Ly-$\alpha$ flux appears to be emanating from A. The lack of metal lines in A's spectrum suggests the presence of a metal--free region, indicative of either a Pop III star cluster or a DCBH environment.

In this study, we model the formation and evolution of the CR7 system in the context of the standard theory of structure formation in a $\Lambda$CDM Universe, and we compare the observations of this system with the predictions derived from this model.   
Consistent with the WMAP 7 results \citep{Komatsu:2011p409}, we assume a $\Lambda$CDM cosmology with $\Omega_0=0.265$, $\Omega_b=0.044$, $\Omega_\Lambda=0.735$, $h=0.71$ and $\sigma_8=0.801$.
We discuss the conditions for DCBH formation in Sec.~\ref{sec.conditions}, followed by arguments for the specific case of the CR7 system in Sec.~\ref{sec.cr7}. Finally, our conclusions and discussion are presented in Sec.~\ref{sec.conclusions}.

\begin{figure*}
\centering
\includegraphics[width=1.3\columnwidth,trim={.5cm 13.5cm 4.5cm .25cm},clip]{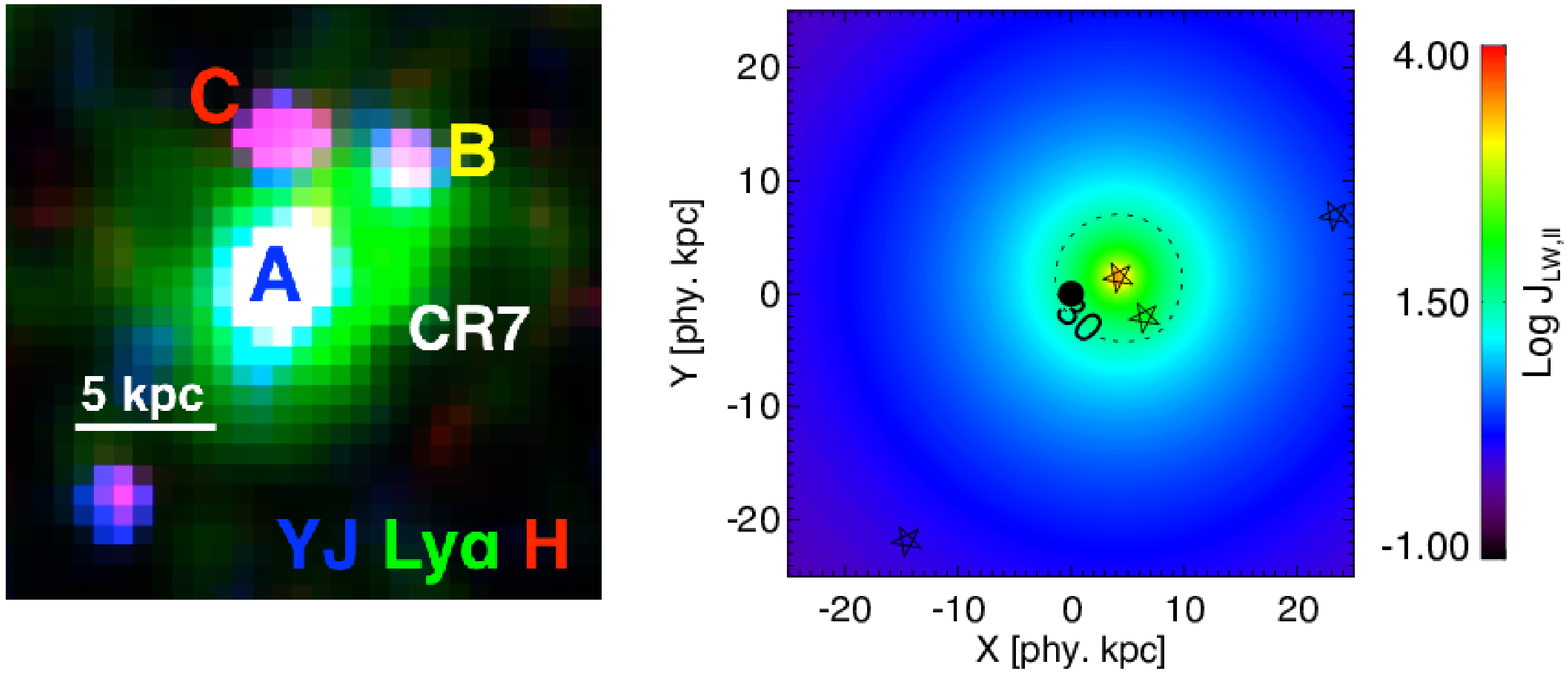}
\includegraphics[width=0.7\columnwidth,trim={-1cm -5cm 0cm 0cm},clip]{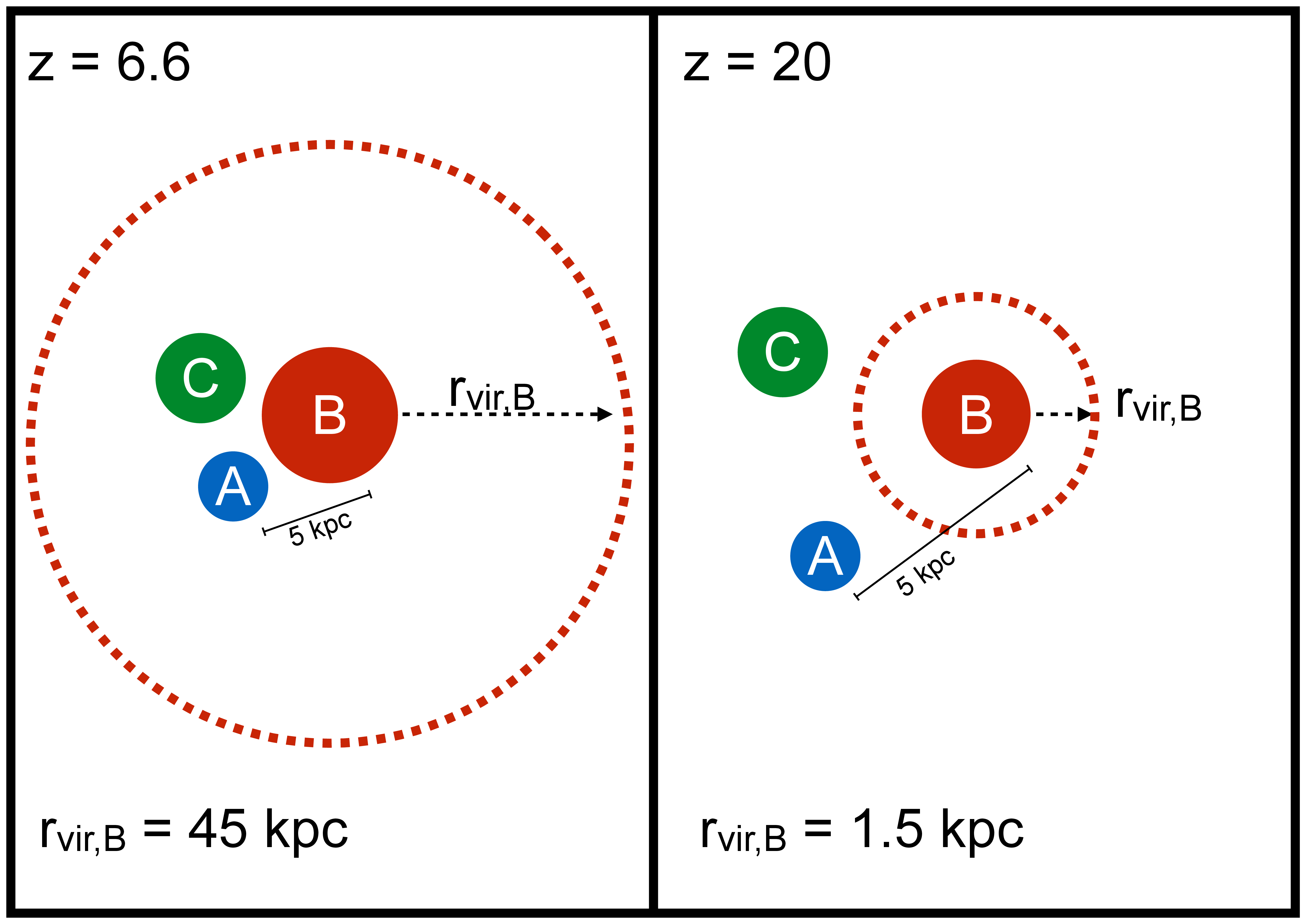}
\caption{The CR7 system. {\textit{Left: Observations. }} A false colour composite of CR7 from S15 constructed using NB921/Suprime-cam imaging along with F110W (YJ) and F160W (H) filters from HST/WFC3. The figure demonstrates the extreme blue nature of component A as compared to components B and C that are much redder. 
{\textit{Middle: Simulation Analogue. }} The LW radiation contour projection in the x--y plane for a DCBH candidate (black) seen in \citet{Agarwal14} with a similar arrangement of neighbouring galaxies (star symbols) as CR7. 
{\textit{Right: Evolution (not to scale). }}The two panels, at $z=6.6$ (${\it left}$) and $20$  (${\it right}$), show the evolution of the virial radius of CR7's host halo, which has a mass of $\sim$ 10$^{12} \msun$  at $z$ = 6.6.  The stellar radiation required for DCBH formation in source A is produced by source B at $z$ $\sim$ 20.  Source A, powered by accretion onto a DCBH, later merges with the larger halo hosting source B.}
\label{fig.cartoon}
\end{figure*}
\section{Cosmic conditions required for DCBH}
\label{sec.conditions}

{The baryonic component of haloes in the first billion years of our Universe's evolution is primarily composed of hydrogen and associated species, e.g. H$_2$, HD, H etc. \citep[e.g.][]{Lepp:1984p3301}, which serve as the primary coolant during the process of gas collapse.  Molecular hydrogen, which serves as a primary coolant in minihaloes,\footnote{$2000 \leq T_{vir}<10^4\ \rm K$ with a corresponding $M_{vir} \sim 10^5  - 10^6 \msun$ which is a function of redshift \citep{Barkana:2001p60}}} can cool gas down to $\sim 200$~K, which at a density of $n=10^5 \ \rm cm^{-3}$, corresponds to a Jeans mass of $\sim 100\msun$, i.e. the characteristic mass for fragmentation. Thus the first metal free stars, or Pop III stars, are expected to form with {with $M_{*} \sim 10 -1000 \msun$ \citep{Bromm:1999p2573,Yoshida:2006p36,Hirano:2014p3818}} The formation of a direct collapse black hole (DCBH) requires pristine gas in an atomic cooling halo predominantly composed of atomic hydrogen \citep[and see references therein]{Volonteri:2010p30, Natarajan:2011p90,Haiman2013review} instead of molecular hydrogen. Thus once the collapse begins, the primary cooling agent is atomic H, which allows the gas to cool isothermally down to a temperature floor of 8000 K.  At a density of $n=10^5 \ \rm cm^{-3}$, this corresponds to a Jeans mass of $\sim 10^5\msun$, which can then undergo a \textit{runaway} collapse into a massive DCBH of $10^4-10^5 \msun$ \citep{Lodato:2007p869}.  We now describe the requirements for this scenario in more detail. 

\subsection{Lyman-Werner radiation}
In order to prevent Pop III star formation in a pristine halo, cooling by molecular H needs to be suppressed and this occurs via  the reaction $\rm H_2 + \gamma_{LW}\rightarrow H + H$, where $\gamma_{LW}$ is a photon in the Lyman--Werner band of 11.2 -13.6 eV. {A \textit{critical} level of LW radiation has been quoted as the requirement to rid the gas of H$_2$, which leads to a high atomic H fraction in the gas. This ensures that at the onset of collapse, cooling by H$_2$ is subcritical.}
This value of critical LW radiation is often quoted as a specific intensity in units of $10^{-21}\ \rm erg/s/cm^2/Hz/sr$ and has been studied extensively in literature for black--body and power--law type spectra. The assumption made in these studies is that Pop II stars can be represented by a T = $10^4$~K black--body spectrum, and Pop III stars by a T = $10^5$~K black--body spectrum, which leads to a J$_{crit} \sim 30 -100$ from Pop II, and J$_{crit}\sim 10^{3-4}$ from a Pop III type stellar population \citep{Omukai:2001p128,Shang:2010p33,WG11}.

Much progress has been made recently with regards to the determination of J$_{crit}$. For instance, implementing a more self-consistent and updated chemical framework relevant to the collapse of pristine gas, can lead to a factor of $\sim$~few difference in the determination of J$_{crit}$ \citep{Glover2015a,Glover2015b}. Also, using realistic spectral energy distributions (SED) to model the Pop III/PopII stellar populations can lead to a significant difference in the qualitative interpretation of J$_{crit}$ \citep[A15a and A15b hereafter]{Sugimura:2014p3946,Sugimura:2015ut, Agarwal15a, Agarwal15b}. 
In their study, A15b have shown that, depending on the detailed properties of the stellar population, the value of J$_{crit}$ can vary over even up to 2 orders of magnitude. This is due to the fact that the reaction rates for the dissociation of H$^-$ and H$_2$, k$_{de}$ and k$_{di}$, govern the fate of collapsing gas. These rates depend on the shape of the SED (A15a) and the net output of photons in the $\sim$ 1eV and LW band respectively. Thus in order to understand if molecular cooling is suppressed in a pristine halo, one must analyse the SED used to model the stellar population(s) and compute the k$_{de}$--k$_{di}$ values to check for DCBH formation (A15a, A15b).

\subsection{Fragmentation and formation of the seed}

Once cooling by molecular hydrogen is suppressed and the Jeans mass has been reached, the gas cloud must withstand fragmentation into stars and lose its angular momentum in order to form a dense core that will ultimately result in the formation of a DCBH. Several authors have studied this in idealised hydrodynamical simulations of isolated haloes, where the gas is allowed to cool and collapse to very high densities ($n \sim 10^{17-18}\rm \ cm^{-3}$) via the atomic cooling channel \citep[e.g.][]{Latif:2013p3629,Schleicher:2013p3661,2016MNRAS.456..500S}. Both turbulence and disc formation have been reported as excellent agents that can rid the gas of its angular momentum resulting in formation of a dense core fed by inflows with accretion rates $\sim 0.1-1 \ \msun/yr$ \citep{Johnson:2011p704,2015MNRAS.450.4411C}. The core can then either form a supermassive star $\sim 10^{5} \msun$ \citep{Hosokawa:2013p3513,Schleicher:2013p3661}, which undergo collapse before becoming thermally relaxed and fully convective, resulting in a $10^{4-5} \msun$ DCBH, or a quasi--star where a dense core is embedded in a optically thick cloud of gas, where radiative losses are inefficient, thus the accretion process becomes super--Eddington for the system overall resulting in a DCBH \citep[e.g.][]{Begelman:2008p672,Spaans:2006p58}.
\subsection{Halo Growth history}
The gas in the halo that hosts the DCBH must be free of metals, which implies no previous in--situ star formation, as well as no pollution from nearby galaxies through stellar winds and/or supernovae. Prior in--situ star formation can be prevented in a halo by a moderate external LW specific intensity, J$_{LW}$, if virial temperatures range between $2000<T_{vir}<10^4$~K (minihalo). The necessary conditions for a halo to host a DCBH are listed below.

\begin{enumerate}
\item {While its virial temperature is $\rm 2000 \leq T_{vir} < 10^4 \ K$, the halo must be in close proximity ($< 20$~kpc) of a galaxy (or galaxies) which produce a large enough LW flux to shut down Pop III star formation \citep[A15b]{Agarwal12,Agarwal14}.}
\item {When the halo has grown to $\rm T_{vir} = 10^4 \ K$, it must be exposed to a large enough LW flux for H$_{\rm 2}$ formation to be supressed, resulting in DCBH formation.}
\item {The halo must be metal free throughout (i) and (ii) above.}
\end{enumerate}

Having outlined the conditions under which a DCBH can form in the early universe, we now examine whether or not CR7 is a viable host for such a  massive seed BH.

\section{The CR7 System}
\label{sec.cr7}
CR7 is an excellent candidate for a Pop III cluster due to the absence of metal lines and the redshift at which it has been observed. However this explanation has been questioned by P15 on the basis of their simulations where they do not find any Pop III clusters that have the right Ly$\alpha$/HeII line luminosities, unless there is a burst of $10^7 \msun$ of young Pop III stars with ages $<5$~Myr. Also the possibility that CR7 could be harbouring an AGN or Wolf--Rayet stars has been shown to be inconsistent with the much broader full width half maxima (FWHM) expected for such systems (S15).
Here we argue that in the system CR7, source A is an excellent candidate for a DCBH, while sources B \& C are the evolved stellar systems that have enabled the formation of a DCBH in it through their LW radiation field. The spectro--photometric observations of CR7 reveal that it is composed of 3 distinct components, A,B \& C, seperated by projected distances of $\approx 5$~kpc (M15, S15). {A description of the system is shown in Fig.~\ref{fig.cartoon}. We highlight the observations from S15 in the left panel with the false--colour composite, Fig. 7 of their study. The LW radiation contour projection of the DCBH candidate, DC4, in the x--y plane taken from the simulation presented in \citet{Agarwal14} is shown in the middle panel. DC4 satisfies all the conditions outlined in Sec.~\ref{sec.conditions} and exhibits a similar configuration as CR7 with its neghbouring galaxies.\footnote{Note that this is for illustration purposes only, as the box--size and high--resolution of the simulation in \citet{Agarwal14} chosen to accurately capture low mass minihaloes ($10^{5-6} \msun$) does not allow for the collapse of a $10^{12} \msun$ halo at $z=6$.}  Finally the size evolution of component B with redshift, and A's assumed distance from it are shown in the right panel.}

\subsection {Description of the system}

CR7 is the brightest Ly$\alpha$ emitter at $z>6$ to date with $L_{Ly\alpha} \sim 10^{43.93} erg/s$ and $L_{HeII} \sim 10^{43.26} erg/s$, with no metal lines detected. The photometric observations from the \textit{Wide Field Camera 3} (F110W: YJ filter; F160W:H filter) suggest that A is extremely \textit{blue}, while B \& C are \textit{redder} in comparison. The FWHM for Ly$\alpha \sim 266 \ km/s$ and HeII $\sim 130\ km/s$, with corresponding equivalent widths (EW) of $\sim 80\ \rm \AA$ and $\sim 230\ \rm \AA$.

We re--reduced and re--analyzed the Spitzer/IRAC imaging data over CR7. In particular, we use our photometry tool MOPHONGO \citep[e.g.][]{Labbe2010b,Labbe2010a,Labbe13} to derive accurate IRAC fluxes for the three different components of CR7. We model these components based on the WFC3/IR images and derive their IRAC fluxes by degrading the HST images to the IRAC PSF and fitting deriving the appropriate flux normalization for each \citep[e.g.][]{Labbe2015}. Doing this we find that $\sim 70 \%$ of the $3.6 \mu m$ and $4.5 \mu m$ flux emanates from A, with sources B \& C contributing only 30 \%, which is in contrast to the findings of S15.\footnote{{The importance of accurate Spitzer/IRAC flux derivations will be highlighted in an upcoming study (Agarwal et al. 2016 in prep).}} As we will see in the following sections, this is critical to disentangle the SED modelling of components A, and B \& C. 

\subsection{SED Fitting: B \& C}

\begin{figure}
\centering
\includegraphics[width=0.75\columnwidth,angle=90,trim={0cm 1cm 1cm 1cm},clip]{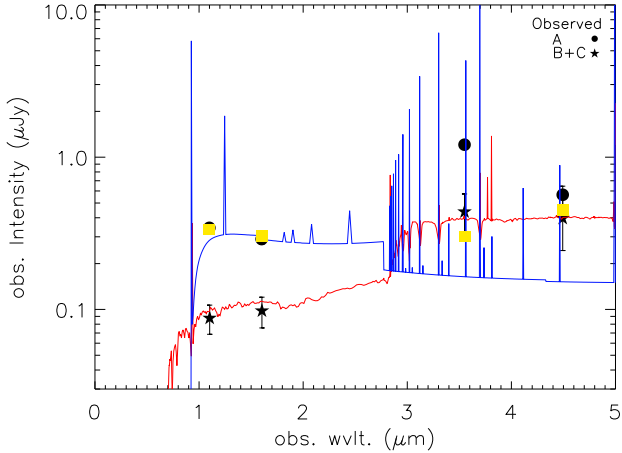}
\caption{Here we compare our SEDs to the observations, assuming the source stellar populations described in Section 3.3 and the BH accretion disk spectrum described in Section 3.8. The solid circles represent the observed flux for source A, whereas the star symbols are the observations for B \& C. We fit source A (blue) with a multi--colour accretion disk spectrum of a BH (along with its nebular component), whereas sources B \& C is fit (red) with an evolved stellar population that is $\sim 700$ Myr old with a metallicity of $0.04 \zsun$ (see text for more details). We also plot in yellow squares, the resulting broadband fluxes in the corresponding band from our best fit model for component A. }
\label{fig.SED}
\end{figure}

{In their study, S15 explored the Pop II explanation of CR7 by trying to fit its spectro--photometric data with a stellar population of $Z=0.2 \zsun$ and age of $\sim 800$  Myr, amongst other cases. 
Using their parameter space as a prior, we proceed} to fit the spectro--photometric data of sources B \& C as two distinct stellar components. Our best fit shown in Fig.~\ref{fig.SED} (red curve) is obtained when:
\begin{itemize}
\item B is fit with a 700 Myr old stellar population, with an exponentially decreasing \footnote{we use here an e--folding time of 200 Myr} SFR from $z\approx 23 - 6.6$, such that at $z=6.6$ it has a SFR of $\sim 2 \msun/\rm yr$ and $M_* = 2 \times 10^{10} \msun$
\item C is fit with a 300 Myr old stellar population, with an exponentially decreasing SFR starting at $z\approx9 - 6$, such that at $z=6.6$ it has a SFR of $\sim 1 \msun/\rm yr$ and $M_* = 7 \times 10^{8} \msun$
\end{itemize}

We use our own SED modeling without any dust attenuation, based on the {\textit {Yggdrasil}} code \citep{Zackrisson11} assuming a stellar metallicity of $0.04 \zsun$, and accounting for nebular emission with the same metallicity. Despite the known degeneracies in SED modelling, we were unable to fit sources B \& C with a single stellar population, and/or assuming exponentially increasing SFR. Our model suggests that bulk of the star formation in B \& C occurred at $z>>6.6$, proving to be viable sources of LW radiation for direct collapse in source A. {In principle C should have a lower metallicity than B owing to its age, however, lowering the metallicity of C only marginally changes the nature of the fit. This can also be attributed to the degeneracies in the fitting parameters of the two populations.}

\subsection{Lyman-Werner history}
\begin{figure}
\centering
\includegraphics[width=0.75\columnwidth,angle=90,trim={0cm 0.57cm 1cm 1cm},clip]{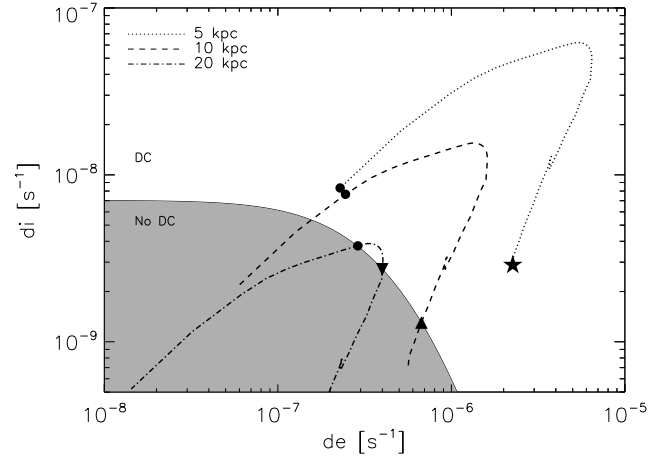}
\caption{Tracing LW history of CR7. The photo--detachment rate ($k_{de}$) and the photo--dissociation rate ($k_{di}$) computed at a distance of 5 (dotted line), 10 (dashed line) and 20  kpc (dashed-dotted line) at each redshift from the stellar system B \& C, plotted against the curve that demarcates the region in the $k_{de}-k_{di}$ parameter space where DCBH formation is permissible. The $k_{de}-k_{di}$ {\textit tracks} that allow for DCBH formation all begin at $z\sim23$ (solid black circles) but end at different times due to the evolution of the SED and the separation assumed between A and B\& C. The upside down triangle, upright triangle and the star symbol correspond to $z\sim16, 7.2, 6.6$.}
\label{fig.lw1}
\end{figure}

Having obtained the SEDs and the star formation history for the stellar systems B \& C, we then compute it's LW history all the way back to the first instance when B \& C were forming stars. Following the analysis of A15a and A15b, we compute the photo--dissociation ($k_{di}$) and photo--detachment ($k_{de}$) rates within a 5 kpc sphere around B \& C at each redshift and find that at $7<z<23$, the stellar system B \& C is able to produce the right set of $k_{de}-k_{di}$ values that can lead to DCBH formation in a pristine atomic cooling halo within the 5 kpc sphere. Even if we relax the constraint on the separation and assume a larger sphere to account for their dynamical evolution, we find that B \& C can still produce conditions ($k_{de}-k_{di}$) suitable for DCBH formation out to a distance of 10 kpc (20 kpc) during $8<z<23$ ($19<z<23$). We show the results of this calculation in Fig.~\ref{fig.lw1}, where the solid curve divides the $k_{de}-k_{di}$ parameter space into regions where DC is permissible or not, and the dotted, dashed, dashed--dotted lines correspond to a distance of 5, 10 and 20 kpc from the stellar sources.

\subsection{Metal pollution}
In order to allow DCBH formation in A, sources B \& C must not pollute its site with metals.
No metal lines are detected in component A (S15), suggesting that metal pollution has not occurred. Despite this, taking a conservative approach, we estimate the range in redshift when A could have been first polluted by sources B \& C and find that with a typical wind speed of $100\rm \ km/s$, it would take the metals 50 Myr to reach A at a separation of 5 kpc. Of course a larger separation would only go in the direction of increasing the time it takes for the metals to reach A. Therefore we assume as a conservative pessimistic case, that the separation is 5 kpc. 
This provides an estimate of the latest redshift by which A must form a DCBH which is at $z\approx 19$. Thus the window for DCBH formation in A is $19<z<23$.\footnote{We note that, even if metals did reach source A, it is possible that they did not mix thoroughly with the dense gas from which the observed nebular emission originates \citep[see e.g][]{Cen:2008p841}.} {Following the release of our work, \citet{TilmanCR7} independently analysed the nature of CR7 in a cosmological framework. An extensive analysis of the metal pollution history of A can be found in their study, which is in excellent agreement with us  and favours the DCBH explanation of CR7 as well.}

\subsection{Source A as a DCBH}
Following the argument in the sections above, it is clear that source A could have formed a DCBH between $19<z<23$. Assuming that a DCBH with a seed mass of $10^5 \msun$ ($10^4$) did form in A at $z=21$, it can grow to $M_{BH} = 7.5\times10^{10} \msun$ ($7.5\times10^9$) by $z=6.6$ via Eddington accretion with $f_{edd}=1, \epsilon = 0.1$.  
To better understand the final state of the BH in source A, we model its SED as a multi--colour accretion disk surrounded by photoionized gas. 

The nebular emission produced by the latter is predicted as in \citet{Zackrisson11}, using the photoionization model Cloudy \citep{Ferland98} under the assumption of metal-free, constant-density ($100\rm \ cm^{-3}$), radiation--bounded nebula. The total SED is generated by adding the nebular SED to that of the accretion disk, correcting the Ly$\alpha$ line by a further escape fraction ($f_\mathrm{esc,Ly\alpha}$) to accommodate possible flux losses in the IGM while setting the flux at wavelengths shortward of Ly$\alpha$ to zero to simulate the Gunn-Peterson trough. The integrated HST (NIC3/F110W, NIC3/F160W) and Spitzer (IRAC/3.6$\mu$m: IRAC Ch1 IRAC, IRAC/4.5$\mu$m: IRAC Ch1) broadband fluxes are then computed from this final SED redshifted to $z=6.6$ and corrected for the luminosity distance. 

Our fit to the observables is shown in Fig.~\ref{fig.SED} (blue), using the model described above with a $M_{BH}\sim 4.4\times 10^6 \msun$, accreting at $f_{Edd}=0.4$, with a $f_{esc,ion}=0$ and $f_{esc,Ly\alpha}=0.16$. Note that this is the first study that attempts to fit a single spectrum to the proposed accreting BH powering clump A. We find that the Ly$\alpha$ and HeII line fluxes, the HeII equivalent width and the IRAC Ch2 flux are well reproduced, i.e. L$_{Ly\alpha}\sim 9\times 10^{43}\ erg/s$, L$_{HeII}\sim 2.1\times 10^{43} erg/s$, $\frac{L_{HeII}}{L_{Lya}}\sim 0.23$, EW(HeII)$\sim 85 \ \rm \AA$. Our model falls short in the IRAC Ch1\footnote{We emphasize, therefore, that further modeling of the nebular emission and source spectrum is warranted.}, but we point out that in this study we fit the EW of the HeII line, the Ly$\alpha$ luminosity, the $\frac{\rm Ly\alpha}{\rm HeII}$ line ratio, and the HST \& Spitzer fluxes simultaneously with a single accreting BH model. This ensures that there is no longer a degeneracy in the $M_{BH}-\dot{M}_{BH}$ parameter space since only specific values of the black hole mass and accretion rate give rise to the right slope (and normalisation) for the multi--colour disk spectrum, which leads us to the re--emitted nebular component.

The final BH mass and accretion rate is also consistent with the formation epoch of $z\sim 20$, if we assume a seed mass of $2\times 10^4 \msun$ (typical range for DCBH), accreting constantly at 40\% of the Eddington rate. A range of seed masses, formation epochs ($19<z<23$) and accretion rates can be accommodated by invoking a time varying accretion rate as long as we end up at $z=6.6$, with a final $M_{BH}\sim 4.4\times 10^6 \msun$ and $f_{Edd} \sim 0.4$.

\subsection{Putative halo assembly histories}
Now let us examine the assembly history of the halo that hosts source
A of the CR7 system over redshift to assess its feasibility as a DCBH
site.  A stellar mass of $\sim10^{10}$ at $z=6.6$ for system B implies
a DM host halo mass of the order of $10^{12} \msun$
\citep{Behroozi2013}. Such a DM halo has a virial radius of
$\sim 45$~kpc (physical), and it is thus safe to assume that the
entire CR7 system is embedded within the virial radius of B's host
halo at $z=6.6$. 

\begin{figure}
\centering
\includegraphics[width=0.75\columnwidth,angle=90,trim={0cm 0.8cm 1cm 1cm},clip]{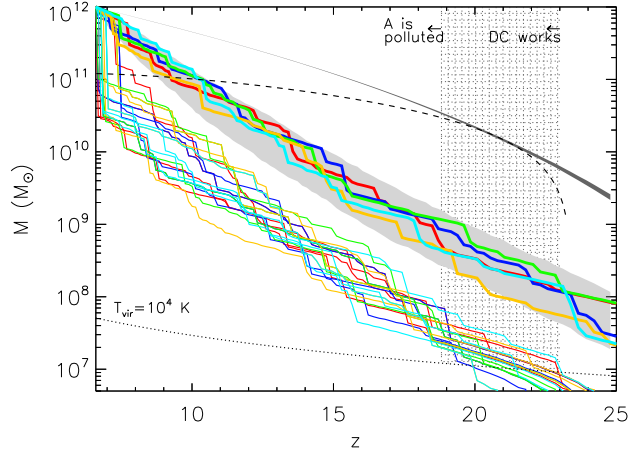}
\caption{Mass assembly history of CR7. We associate a $10^{12} \msun$ DM host halo at $z=6.6$ to the stellar system B. 
The light grey region bounds the 16$^{th}$--84$^{th}$ percentile of the 500 realisations employed to compute the mass assembly history of B's host halo. 
The 5 thick lines are just few of the such realisations colour coded in order to be distinguishable. The thin lines (possible host haloes of A), are the mass assembly histories of what end up being subhaloes of the host haloes (thick lines of corresponding colours) of B. The hatched region marks the time--window most suitable for DCBH formation. The vertical jump in the thin lines at $z\ltsim7$ indicates accretion into their current $z=6.6$ hosts, i.e. the transition from being host haloes to subhaloes. The atomic cooling limit  $T_{vir}=10^4\ K$ is plotted as the dotted line. {An explanation of the dark grey shaded region and the dashed line can be found in the Appendix.}}
\label{fig.dmh}
\end{figure}

Using the method of \citet{2008MNRAS.383..557P}, and adopting the WMAP7 cosmology outlined earlier, 
we construct an ensemble of 500 merger trees for a host
halo of mass $10^{12} \msun$ at $z=6.6$. These are used to construct
the average mass accretion history (MAH), and its halo-to-halo
variance, using the method outlined in \citet{VDB2014}.  The
light grey region in Fig.~\ref{fig.dmh} bounds the
16$^{th}$--84$^{th}$ percentiles, while the solid, thick lines
indicate a random subset of the 500 realizations used.

Next we search each merger tree for branches (corresponding to dark
matter subhaloes) that could host a DCBH (i.e., source A).  The dark
grey region in Fig.~\ref{fig.dmh} bounds the time frame most suitable
for DCBH formation, i.e. the LW conditions are met at $z<23$ and
metals from B \& C have not yet made it to A at $z>19$. The dotted
line marks the $T_{vir}=10^4$~K limit, below which baryonic gas cannot
cool. Hence, in order for a DCBH to form, we require that its host
halo (which is to become a subhalo of the host halo of B at a later time) upcrosses
the atomic cooling limit in the redshift range $19 < z_{upcross} <
23$.  Also for this halo to form a DCBH that is sufficiently
massive it needs to be able to accrete enough gas so that it can grow
from its seed mass (here assumed to be $M_{BH}^{seed} = 10^5 \msun$)
to its inferred mass of $4.4 \times 10^{6} \msun$ at $z=6.6$. If we
make the conservative assumption that the BH growth is such that it
ends up on the $z=0$ BH mass - bulge mass relation, roughly $10^{-3}$ of the baryonic mass accreted by the DCBH's host
halo \citep{Li:2007p54,Kulier14} has to be funnelled into the black hole. Taking into account that the
mass of a dark matter halo is arrested once it becomes a subhalo, due
to the tidal forces of the host halo \citep[see][and references therein]{Arthur2015}, this puts the following criterion on the mass
history of the halo that hosts the DCBH (source A):
\begin{multline*}
\nonumber M_{DM}(z_{acc}) - M_{DM}(z_{upcross}) > \\
{M_{BH}(z=6.6) - B_{BH}^{seed}(z_{upcross}) \over 10^{-3} f_b} 
\simeq 
2.9 \times 10^{10} \msun
\end{multline*}
Here $z_{acc}$ is the redshift at which the halo that hosts the DCBH
is accreted into the host halo of source B, and $f_b \simeq 0.15$ is
the universal baryon fraction. Finally, in order for the BH to be
observable at $z=6.6$, it must still be accreting at that redshift.
The growth of the halo A is arrested once it becomes the subhalo of B, thus the 
BH in A grows from the gas reservoir that was accumulated prior to A's accretion by B.
Under the assumption that gas accretion onto the BH is depleted 
within 1 free fall time and that we require that A should 
still be accreting, we deduce that the merger of A onto B 
occurred between z=6.6. and z=7.5.

The thin lines in Fig.~\ref{fig.dmh} are a random subset of mass
accretion histories of subhaloes of host haloes that reach a mass
of $10^{12} \msun$ at $z=6.6$ that meet the following criteria:
\begin{itemize}
\item upcross the atomic cooling limit between $19 < z_{upcross} < 23$,
\item accrete onto the host halo of source B at $6.6 < z_{acc} < 7.5$,
\item have enough gas to feed the BH, i.e. $M_{DM}(z_{acc}) -
 M_{DM}(z_{upcross}) > 2.9 \times 10^{10} \msun$,
\end{itemize}
and that therefore potentially host a DCBH. We find that a host
halo of $10^{12}\msun$ at $z=6.6$ has on average 0.2 subhalos that
meet all these criteria. Out of 500 realizations, 78 ($\sim
16\%$) have one subhalo that meets all three criteria listed above,
while 1.5\% of the host halos have 2 subhaloes or more that meet all
criteria.We emphazize that the criteria used above are fairly
conservative: in principle BHs at high redshift could accrete more
than 0.1 percent of all the baryons \citep{Alexander14,Volonteri:2005p793}, 
and it might take significantly longer
than one free-fall time for the gas to be accreted by the BH after
being accreted by its halo. For example, if we assume that 1,
 rather than 0.1, percent of the baryons are accreted by the BH, we
 find that each host halo has on average 3 subhalos that meet all
 criteria. Of all the realisations, 96\% haloes have at least one subhalo that meet all the criteria and 79\% have 2 or more such subhaloes.
 
 Hence, we conclude that a typical host halo of mass
$10^{12}\msun$ at $z=6.6$ can easily host one or more subhalos with a
DCBH, {\it as long as it has been exposed to sufficient LW
 radiation}. In the case of CR7, this implies that the subhalo that
hosts source A should have remained within a radius of roughly 20 kpc
(physical) from source B (and/or C) over the redshift range $19 < z <
23$ (which is the epoch at which we postulate that the DCBH
formed). We address this issue in the next subsection.


\subsection{Separation between the components}
\label{sec.sep}
We note here that we computed the LW conditions from B \& C, assuming
that the separation between the stellar system and A remains constant
in time. This is clearly an oversimplification. In reality, prior to
being accreted into the host halo of B, the physical distance between
the halos hosting A and B first increases due to the Hubble expansion,
followed by a decrease as A is starting to fall towards B. After
accretion, the subhalo that hosts A will orbit B. A $10^{12} \msun$ DM
halo at $z=6.6$ has a virial radius of $r_{vir} \simeq 45$~kpc
(physical). According to the spherical collapse assumption, the matter
that has collapsed to form this halo (and its substructure),
originated from a Lagrangian volume with a radius that is twice that
of the virial radius. This Lagrangian volume has a physical radius of
$\sim 30$~kpc at $z=20$, which is approximately the redshift at which
we postulate that the DCBH formed. If we assume a roughly constant
density within that volume, there is roughly a 30 percent probability
that a mass element within that volume is located within 20 kpc from
the center where source B is postulated to reside. For comparison,
according to the mass accretion histories, the typical main progenitor
of a host of mass $10^{12} \msun$ at $z=6.6$ has a mass of $\sim 10^9
\msun$ at $z=20$, which corresponds to a virial radius of only $\sim
1.5$~kpc (physical). According to this (very rough) calculation, there
is a $\sim$1 in 3 chance that the subhalo hosting A was located
within about 20 kpc from B at the crucial redshift interval $19-23$.

Based on these estimates, the probability that a host halo of mass
$10^{12} \msun$ at $z=6.6$, whose central galaxy has the same LW
history as source B, hosts {\it at least} one subhalo that could have
formed a BH mass similar to that of source A is given by
\begin{equation}
P = 1 - \sum_{N=0}^{\infty} P(N) \, \left({2 \over 3}\right)^N
\end{equation}
where $P(N)$ is the probability that the host halo contains $N$ subhaloes
that meet the three criteria listed in the previous subsection.
Using the probabilities inferred from our merger trees, we infer
that $P \sim 0.06$ for the case where $0.1\%$ of the baryons are allowed to be accreted by the BH, 
whereas $P \sim 0.63$  for the case where $1\%$ of the baryons are allowed to be accreted by the BH. Thus it is not at all unlikely that source A may indeed have formed a DCBH.

{Note that eq. 1 is a fairly conservative lower limit; if we were to assume that the density distribution within the 30 kpc Lagrangian volume is centrally concentrated rather than uniform, the probability to host at least one DCBH increases. We further include a discussion on the temporal-spatial distribution of the stellar mass in B, and its impact on the LW history of the system in the Appendix.}

\section{Discussion and Conclusions}
 \label{sec.conclusions}
 
The discovery of the brightest Ly$\alpha$ emitter in the COSMOS field \citep{Matthee15a}, CR7, has generated active interest about its origin (S15,P15). The possibility that it could be a Pop III galaxy or an obese black hole galaxy (OBG) \citep{Agarwal13} has been suggested by S15.
Recently, P15 have reported that it is unlikely that component A in CR7 could be harbouring a purely Pop III population. This they argue is because to produce the Ly$\alpha$ \& HeII line luminosities, a combination of an extremely top--heavy IMF with a $\sim 10^7 \msun$ burst of very young ($t_*<5$~Myr) Pop III stars is required. Furthermore they  claim that component A could potentially be an DCBH, with a seed mass $\sim 10^5 \msun$ and growth period lasting 120 Myr. 
However, they do not explore the formation history of the black hole, or the MAH of the DM host halo for this system.

Here we do all of the following to build a comprehensive history for the stellar components, the DCBH and the MAH of the DM haloes simultaneously:

\begin{itemize}
\item model the SED of the stellar components B \& C
\item calculate the LW history from B \& C incident on A
\item construct the MAH of the DM host haloes of CR7 down to $z=6.6$
\item estimate the window of possible metal pollution of A by B \& C
\item constrain the epoch of DCBH formation taking all of the above into account
\end{itemize}

Including these multiple constraints we find component A satisfies all the pre--requisites for hosting a DCBH that has now evolved to $\approx 4.4 \times 10^6 \msun$ at $z=6.6$, accreting at $f_{edd} \approx 0.4$.
Our model for the accreting DCBH's spectrum (along with its nebular re--emission) successfully reproduces the observed Ly$\alpha$ and HeII line flux, the HeII equivalent width, and the HST and Spitzer fluxes, {but falls short in the IRAC Ch1 band, warranting further modelling of the nebular component.}
 
With observations of this system at $z=6.6$, discriminating between the high--z seeding models for example Pop III remnant seeds vs DCBH, is in principle difficult. However, we show in this study that the expected mass assembly histories of the DM host haloes, and the evolution of  the LW radiation strongly support the case for a DCBH assembling between $19<z<23$ in A.

Detecting higher redshift counterparts of such systems would shed more light on the detailed assembly history of the seed BH. Such observations, particularly if the seed grows via super--Eddington accretion for a short period of time \citep[P15]{FabioSED} might be accessible with future X--ray missions like the \textit{Athena} or \textit{X--ray surveyor}. Deep X--ray or ALMA observations of the CR7 system could provide additional constraints to help pin down the DCBH growth history.
 
\section{Acknowledgements}
{The authors would like to thank the anonymous referee whose suggestions greatly improved the paper}. The authors would also like to thank David Sobral and Jorryt Matthee for their useful comments on the manuscript. BA would like to thank Pascal Oesch for his inputs that greatly helped in shaping the manuscript. BA would also like to thank Laura Morselli, Chervin Laporte and Jonny Elliott for their help during the preparation of this study. PN acknowledges support from a NASA-NSF Theoretical and Computational Astrophysics Networks award number 1332858. BA acknowledges support of a TCAN postdoctoral fellowship at Yale. E.Z. acknowledges research funding from the Swedish Research Council (project 2011-5349).  Work at LANL was done under the auspices of the National Nuclear Security Administration of the US Department of Energy at Los Alamos National Laboratory under Contract No. DE-AC52-06NA25396.

\bibliographystyle{mn2e}
\bibliography{babib}

\section*{{Appendix}}
\subsection*{Spatial--temporal distribution of B's stellar component}
In Fig.~\ref{fig.dmh}, the slim--dark--grey region shows the 16$^{th}$--84$^{th}$ percentile distribution of all progenitors of B above the atomic cooling limit at that redshift. The grey dashed line represents the total stellar mass in B divided by the baryonic mass fraction. This implies that B can host the stellar mass we associate with it at $z=6.6$ if its progenitors convert 100\% of their baryons into stars during $z=21-19$. Such an efficient baryon to star conversion is highly disfavoured at lower redshifts, but one can not rule out such high efficiencies in the first billion years of galaxy formation. 
{However, even if B's progenitors convert 6.25\% of their baryons into stars, the DCBH formation window of $z=21-19$ is still maintained. This can be understood from Fig.~\ref{fig.lw1}, where the 20 kpc separation still leads to a DCBH formation window (pre--metal pollution) of $z=21-15$. The curve can instead be interpreted as a $\frac{5^2}{20^2} \approx 6\%$ (scaled by distance squared) lower baryon to stellar conversion factor at the same original separation of 5 kpc.}

Furthermore, distributing the total stellar mass in B equally among all its progenitors, instead of the main progenitor only, challenges our assumption of A being at a constant distance of 5 kpc from the LW source at all times. In the regime $10<z<23$, we re-distributed the stellar mass expected in B at each time--step of our calculation in $n_{prog}$ progentiors, where $n_{prog} = 50-100$. We then computed the average separation between A and such progenitors at any given redshift by randomly distributing the progenitors in B's Lagrangian volume (see Sec.~\ref{sec.sep}).
Assuming a uniform density distribution, we can write

\begin{equation}
\frac{M_B}{\frac{4}{3}\pi r_L^3}=\bar\rho
\end{equation}
where at any given redshift: $M_B$ is B's main progenitor mass, $r_L$ is the physical extent of the Lagrangian region where one would expect to find all of B's progenitors; and $\bar\rho$ is the average cosmic matter density.

After randomly distributing $n_{prog}$ progenitors over a scale length $S$, we compute the quantity

\begin{equation}
R =  \left(\sqrt{\frac{1}{n_{prog}} {\sum_{i=1}^{n_{prog}}\frac{1}{r_i^2}}}\right)^{-1}
\end{equation}
where $r_i$ is the distance of each progenitor from A.

This is because the total flux reaching A, $F_{A}$,  at any given redshift from $n_{prog}$ progenitors is of the form

\begin{equation}
F_A = \sum_{i=1}^{n_{prog}}\frac{L_i}{4\pi r_i^2}
\end{equation}
where $L_i$ is the luminosity of the stellar component in each progenitor such that
\begin{equation}
L_i = \frac{L}{n_{prog}}
\end{equation}
where $L$ is the total luminosity of the stellar component. Thus, the effective average distance of B's progenitors from A can be written as

\begin{equation}
{\rm{r_{ eff}}} = \frac{R}{S}r_{L}
\end{equation}

Between $10<z<23$, we find $10\ \rm{kpc}> r_{eff}>5\ \rm{kpc}$, thus proving that distributing the stellar mass of B in it's progenitors does not affect the LW history of the system. Note that this computation does not depend on the choice of $S$ as long as it is kept constant throughout.
\end{document}